\documentclass{JHEP3} 

\title{Physical wavelets: Lorentz covariant, singularity-free, finite
  energy, zero action, localized solutions to the wave equation}

\author{Matt Visser\thanks{Research supported by the 
Marsden Fund administered by the Royal Society of New Zealand}\\
School of Mathematical and Computing Sciences, 
Victoria University of Wellington, 
New Zealand.
\\
E-mail: \email{matt.visser@vuw.ac.nz}} 

\abstract{Particle physics has for some time made extensive use of
  extended field configuations such as solitons, instantons, and
  sphalerons. However, no direct use has yet been made of the quite
  extensive literature on ``localized wave'' configurations developed
  by the engineering, optics, and mathematics communities. In this
  article I will exhibit a particularly simple ``physical wavelet''
  --- it is a Lorentz covariant classical field configuration that
  lives in physical Minkowski space. The field is everwhere finite and
  nonsingular, and has quadratic falloff in both space and time.  The
  total energy is finite, the total action is zero, and the field
  configuration solves the wave equation. These physical wavelets can
  be constructed for both complex and real scalar fields, and can be
  extended to the Maxwell and Yang-Mills fields in a straightforward
  manner. Since these wavelets are finite energy, they are guaranteed
  to be classically present at finite temperature; since they are zero
  action, they can contribute to the quantum mechanical path integral
  at zero ``cost''.}


\preprint{hep-th/0304081}
\begin{document}
\def\d{{\mathrm{d}}}
\def\L{{\mathcal{L}}}

\section{Motivation}

While the particle physics community has for some time made extensive
use of extended field configurations such as solitons, instantons, and
sphalerons, no direct use has yet been made of the quite extensive
literature on ``localized wave'' configurations developed by the
engineering, optics, and mathematics communities. (For selected
references see~\cite{zed,tippet,LG,LZG,Kaiser,Lekner}.) These
localized waves are classical solutions of the wave equation that are
partially localized in space or time, this localization generally
coming at a cost such as infinite total energy and/or instability
(leading to dispersion or diffraction). The catalogue of known
localized waves is large and growing~\cite{key}, but most of the known
examples are not in a form that would be easy to apply to particle
physics problems.

In this article I will exhibit a particularly simple ``physical
wavelet'' that is more promising from a particle physics standpoint.
It satisfies the properties that:
\begin{itemize}
  
\item It is a localized wave that solves the wave equation.
  
\item It is a Lorentz covariant classical field configuration that
  lives in physical Minkowski space.

\item 
The field is everywhere finite and nonsingular, and has quadratic
falloff in both space and time.
  
\item The total energy is finite, depending on the peak field and the
  width of the pulse.
  
\item The total action is zero.

\end{itemize}
These physical wavelets can be constructed for both complex and real
scalar fields. Extending these idea to the Maxwell and Yang-Mills
fields is straightforward. The simplest case is that of the complex
scalar field and it is to that case that I first turn.

\section{Complex scalar field}

Let $\eta_{ab} = \hbox{diag}[+1,-1,-1,-1]$ be the Minkowski metric
[particle physics signature], let $x_0$ be an arbitrary 4-vector, and
let $\zeta^a$ be arbitrary timelike 4-vector, then
\begin{equation}
\phi(x) = - 
{\phi_0 \; (\eta_{ab} \; \zeta^a \; \zeta^b)
\over
\eta_{ab} \;[x^a-x_0^a-i\zeta^a] \; [x^b-x_0^b-i \zeta^b]
}
\label{E:mother}
\end{equation}
is a Lorentz covariant, finite energy, zero-action solution of the
d'Alembertian wave equation $\Delta \phi=0$.  The ``center'' of the
pulse is at $x_0$ and its ``width'' is $a = \sqrt{\eta_{ab} \; \zeta^a
\; \zeta^b}$.  The field is everywhere finite and in fact
\begin{equation}
|\phi(x)| \leq |\phi_0|.
\end{equation}
To see this, use the fact that $\zeta$ is timelike. Then, using the
manifest Lorentz covariance of the field configuration, we can without
loss of generality first translate $x_0\to 0$, and then go into the
zero-momentum frame where
\begin{equation}
\zeta^a = (a,0,0,0).
\end{equation}
Then the field configuration is
\begin{equation}
\phi(x) = 
- {\phi_0 \; a^2
\over
[t-ia]^2 - x^2-y^2-z^2
}.
\end{equation}
That is
\begin{equation}
\phi(x) = 
{\phi_0 \; a^2
\over
r^2-t^2+a^2+2iat
}.
\end{equation}
Once written in this form it is a straightforward exercise to verify
that the wave equation is satisfied.  To see that the field is
everywhere bounded note
\begin{eqnarray}
|\phi|^2 
&=& 
{|\phi_0|^2 \; a^4 
\over
(r^2-t^2+a^2)^2 + 4 a^2 t^2}
=
{|\phi_0|^2 \; a^4 
\over
(r^2+t^2+a^2)^2 - 4 r^2 t^2} \leq 
\nonumber\\
& \leq  &
{|\phi_0|^2 \; a^4 
\over
(r^2+t^2+a^2)^2 -  (r^2 +t^2)^2}
=
{|\phi_0|^2 \; a^4 
\over
a^4 + 2a^2(r^2+t^2)} 
\leq |\phi_0|^2.
\end{eqnarray}
{From} the penultimate inequality we also derive
\begin{equation}
|\phi|^2 \leq {1\over2} \; |\phi_0|^2 \; { a^2 \over r^2+t^2},
\end{equation}
demonstrating the promised quadratic falloff in both space and time.
Indeed for fixed $t$ the magnitude of the field is maximized when
\begin{equation}
r^2 = \max \{ t^2-a^2, 0 \},
\end{equation}
showing that the configuration disperses to spatial infinity at both
$t\to\pm\infty$.

To calculate the 4-momentum, we remind the reader that the
stress-energy tensor for a massless complex scalar is
\begin{equation}
T_{ab} = 
{1\over2}\left[\phi_a^* \; \phi_b+ \phi_a \; \phi_b^*\right]  
- 
{1\over2} \eta_{ab} |\nabla \phi|^2.
\end{equation}
Then
\begin{eqnarray}
\nabla_a T^{ab} &\equiv& 
{1\over2} \Delta \phi^* \; \nabla_b \phi + 
{1\over2} \nabla^a \phi^* \;\nabla_a \nabla_b \phi +
{1\over2} \Delta \phi \; \nabla_b \phi^* + 
{1\over2} \nabla^a \phi \;\nabla_a \nabla_b \phi^* 
\nonumber
\\
&&\qquad\qquad
-{1\over2}   \nabla^b \phi^* \;\nabla_a \nabla_b \phi^*          
  -{1\over2}  \nabla^a \phi^* \;\nabla_a \nabla_b \phi,
\\
&\equiv &
{1\over2} \left[  
\Delta \phi^* \; \nabla_b \phi +  \Delta \phi \; \nabla_b \phi^*
\right],
\end{eqnarray}
which vanishes by the equations of motion. But this means that
\begin{equation}
P^\mu = \oint T^{ab} \; \d \Sigma_b
\end{equation}
is a conserved quantity, the 4-momentum of the configuration, which is
independent of the particular spacelike hypersurface $\Sigma$ chosen to
do the integration. By simple dimensional analysis
\begin{equation}
P^a = C \; |\phi_0|^2 \zeta^a,
\end{equation}
where $C$ is a dimensionless number to be calculated. [Note that
$\zeta^a$ has the dimensions of a position vector --- a distance.]
The energy density is
\begin{equation}
\rho = {1\over2} \left[ |\partial_t \phi|^2 + |\partial_r \phi|^2 \right],
\end{equation}
and in the zero-momentum frame is easily calculated to be
\begin{equation}
\rho = 
{
2 a^4 |\phi_0|^2 (r^2+t^2+a^2)
\over
(r^2-t^2+a^2+2iat)^2(r^2-t^2+a^2-2iat)^2.
}
\end{equation}
For arbitrary $t$ this integrates to 
\begin{equation}
{\mathcal E} = \oint \d^3 r \rho 
= \int_0^\infty 4\pi\; r^2 \rho = {1\over2} \pi^2 |\phi_0|^2 a.
\end{equation}
(This is independent of $t$ as it should be.)  This is the invariant mass
of the field configuration.  By spherical symmetry, the total momentum
is zero.  Thus, for any timelike $\zeta^a$
\begin{equation}
P^a =  {1\over2} \pi^2 |\phi_0|^2 \; \zeta^a.
\end{equation}
Furthermore the Lagrangian is
\begin{equation}
\L = {1\over2} \left[  |\partial_t \phi|^2 - |\partial_r \phi|^2 \right],
\end{equation}
which evaluates (in the zero-momentum frame) to
\begin{equation}
\L = 
{
2 a^4 |\phi_0|^2 (t^2+a^2-r^2)
\over
(r^2-t^2+a^2+2iat)^2(r^2-t^2+a^2-2iat)^2
}.
\end{equation}
It is easy to check that
\begin{equation}
\oint \d^4 x \; \L = 0,
\end{equation}
so that the configuration is zero action.

In summary, what we have is a Lorentz covariant, singularity-free,
finite energy, zero action, exact localized solution to the
d'Alembertian equation.  In many ways this configuration has more
right to be called an ``instanton'' than do the instantons of QFT;
those instantons live in Euclidean signature. This field configuration
lives in real physical time.

Now the fact that there are finite energy solutions to the wave
equation is not a surprise; that these finite energy solutions can
coalesce, bounce, and disperse without producing field singularities
is more interesting. One way of guessing that the field configuration
in equation (\ref{E:mother}) is worth investigating is the following:
It is easy to convince oneself that in 4 Euclidean dimensions the
solution to Laplace's equation with a delta function source at the
origin is
\begin{equation}
\phi(x) \propto {1\over x^2+y^2+z^2+t^2}.
\end{equation}
Thus in (3+1) Lorentzian dimensions the [singular] solution to the
wave equation with a delta function source at the origin is
\begin{equation}
\phi(x) \propto {1\over x^2+y^2+z^2-t^2}.
\end{equation}
If the source is now moved to a real position $x_0^a$ we have
\begin{equation}
\phi(x) \propto {1\over (x-x_0)^2+(y-y_0)^2+(z-z_0)^2-(t-t_0)^2}.
\end{equation}
which is still a singular field configuration. Finally, move the
source away from physical Minkowski space to the complex position
$x_0^a - i\zeta^a$, then
\begin{equation}
\phi(x) \propto {1\over 
(x-x_0+i\zeta^1)^2+(y-y_0+i\zeta^2)^2+(z-z_0+i\zeta^3)^2-(t-t_0+i\zeta^0)^2},
\end{equation}
which is essentially equation (\ref{E:mother}) above.  This style of
approach has been particularly advocated by Kaiser~\cite{Kaiser}. As
we have just seen, if $\zeta$ is timelike the resulting field
configuration is singularity free.  However, for null and spacelike
$\zeta^a$, while the field is still a solution of the wave equation,
the field is not bounded.  Because of the singularities the energy and
action integrals then diverge. Details are deferred for now and will
be presented in sections \ref{S:null} and \ref{S:spacelike} below.

One should also note that in the optics and engineering literature the
most commonly used notations are not manifestly Lorentz covariant.
Thus it is common to see expressions such as
\begin{equation}
\phi \propto {1\over x^2+y^2+[b_1-i(z+t)]\;[b_2+i(z-t)]}
\label{E:noncovariant}
\end{equation}
(see, for instance, \cite{Lekner}\,) whose Lorentz transformation
properties are less than obvious --- in fact this field configuration
is equivalent to equation (\ref{E:mother}) with the identification
\begin{equation}
\zeta^a = -\left({b_1+b_2\over2},0,0, {b_1-b_2\over2}\right); 
\qquad
||\zeta|| = b_1 b_2;
\end{equation}

\section{Real scalar field}

By taking real and imaginary parts of the complex solution above we
can write down two solutions for the real scalar field. Namely
\begin{equation}
\phi_1 = {\phi_0 \; (\eta_{ab} \; \zeta^a \; \zeta^b) \;
\left\{\eta_{ab} \;[x^a-x_0^a] \; [x^b-x_0^b] - \eta_{ab} \; \zeta^a \; \zeta^b \right\}
\over
(\eta_{ab} \;[x^a-x_0^a] \; [x^b-x_0^b] - \eta_{ab} \; \zeta^a \; \zeta^b  )^2 +
4 (\eta_{ab} \;[x^a-x_0^a] \; \zeta^b)^2};
\end{equation}
\begin{equation}
\phi_2 = {2 \phi_0 \; (\eta_{ab} \; \zeta^a \; \zeta^b) \;
\left\{\eta_{ab} \;[x^a-x_0^a] \; \zeta^b  \right\}
\over
(\eta_{ab} \;[x^a-x_0^a] \; [x^b-x_0^b] - \eta_{ab} \; \zeta^a \; \zeta^b  )^2 +
4 (\eta_{ab} \;[x^a-x_0^a] \; \zeta^b)^2}.
\end{equation}
As previously, we can without loss of generality translate $x_0\to0$
and go to the zero-momentum frame $\zeta^a=(a,0,0,0)$, then
\begin{equation}
\phi_1 = {\phi_0 \; a^2  \;
\left\{ t^2-r^2-a^2 \right\}
\over
(t^2-r^2-a^2)^2 +
4 a^2 t^2};
\end{equation}
\begin{equation}
\phi_2 = {\phi_0 \; a^2  \;
 2 a t
\over
(t^2-r^2-a^2)^2 +
4 a^2 t^2}.
\end{equation}
The stress-energy for a real scalar field simplifies
\begin{equation}
T_{ab} =
\phi_a \; \phi_b
- 
{1\over2} \eta_{ab} |\nabla \phi|^2.
\end{equation}
Then
\begin{eqnarray}
\nabla_a T^{ab} 
&\equiv& 
\Delta \phi \; \nabla_b \phi + 
\nabla^a \phi \;\nabla_a \nabla_b \phi -
\nabla^b \phi \;\nabla_a \nabla_b \phi         
\\
&\equiv &
 \Delta \phi \; \nabla_b \phi,
\end{eqnarray}
which vanishes by the equations of motion. The calculation for the
energy-momentum 4-vector now yields:
\begin{equation}
P^a_1 =  {1\over4} \pi^2 |\phi_0|^2 \; \zeta^a = P^a_2.
\end{equation}
The action integral for both of these field configurations is still
zero. [$\oint \,\d^3 x \,\d t \; {\cal L}(\phi_{1,2}) = 0$.]

\section{Maxwell field}

A similar construction can be performed for the Maxwell field. There
are a number of choices one could make, and I will simply pick one
that leads to a relatively simple field configuration. Start by
picking a timelike 4-velocity $V^a$ and a spacelike unit vector $m^a$
orthogonal to it. Now adopt the ansatz
\begin{equation}
A^a = 
\left\{ V^a \; m^b - m^a \; V^b \right\} \nabla_b \psi.
\end{equation}
Here one is automatically in Lorenz gauge~\cite{Lorenz}, $\nabla_a A^a
= 0$, and the Maxwell equations reduce to the wave equation
$\Delta\psi=0$ for the scalar potential $\psi$. The field
configuration is manifestly Lorentz covariant and we can without loss
of generality go to the inertial frame where
\begin{equation}
V^a=(1,0,0,0); \qquad m^a= (0; \vec m).
\end{equation}
In this inertial frame
\begin{equation}
A_a = (\varphi,\vec A) = 
\left( [\vec m\cdot\vec\nabla] \psi, \vec m\; \dot\psi\right),
\end{equation}
where $\vec m$ is a constant unit vector. We shall soon see that this
is tantamount to working in the zero-momentum frame of the field
configuration.  The electric field is
\begin{equation}
\vec E = - \vec\nabla\left([\vec m\cdot\vec\nabla] \psi\right) 
+ \vec m \; \ddot \psi
=  - \vec\nabla\left([\vec m\cdot\vec\nabla] \psi\right)
+ \vec m \; \nabla^2 \psi
= \vec\nabla\times(\vec\nabla\times[\vec m\; \psi]),
\end{equation}
where we have used the wave equation for $\psi$. The magnetic field is
\begin{equation}
\vec B = \vec\nabla\times(\vec m\; \dot\psi) = 
- \vec m \times \vec \nabla \dot\psi.
\end{equation}
The energy density and momentum flux (Poynting vector) are
\begin{equation}
\rho = {1\over2} [\vec E^2+\vec B^2];
\qquad
\vec S = \vec E \times \vec B.
\end{equation}
To evaluate total energy and momentum a useful integration by parts
(subject to suitable falloff at spatial infinity) is
\begin{eqnarray}
\oint \d^3 x \; (\vec\nabla \times \vec X_1)\cdot(\vec\nabla \times \vec X_2)
&=&
\oint \d^3 x \; (\vec X_1 \times \vec\nabla  )\cdot(\vec\nabla \times \vec X_2)
\\
&=&
\oint \d^3 x \; \vec X_1 \cdot \left[ \vec\nabla \times (\vec\nabla \times \vec X_2) \right]
\\
&=&
\oint \d^3 x \; \vec X_1 \cdot 
\left[ -\vec\nabla^2   \vec X_2 + \vec\nabla ( \vec\nabla\cdot X_2) \right]
\\
&=&
- \oint \d^3 x \; \left\{
\vec X_1 \cdot \vec\nabla^2 \vec X_2 + 
(\vec\nabla\cdot \vec X_1)  ( \vec \nabla \cdot\vec  X_2) \right\}.
\end{eqnarray}
Consequently
\begin{eqnarray}
\oint \d^3 x \; \vec E^2
&=& -
\oint \d^3 x \left\{ [\vec\nabla\times(\vec m\psi)] \nabla^2 [\vec\nabla\times(\vec m\psi)]
\right\}
\\
&=& +  
\oint \d^3 x \left\{
(\vec m\psi) \nabla^2 \nabla^2 (\vec m\psi) + 
( \vec \nabla \cdot[\vec  m\psi])  
\nabla^2 ( \vec \nabla \cdot[\vec  m\psi])
\right\}
\\
&=& +  
\oint \d^3 x \left\{
\psi \nabla^2 \nabla^2 \psi + 
( \vec m \cdot \vec  \nabla\psi)  
\nabla^2 ( \vec m \cdot \vec  \nabla\psi)
\right\}.
\end{eqnarray}
Now let us assume the potential $\psi$ is spherically symmetric
$\psi(r,t)$. Then averaging over angular variables is the same as
averaging over orientations of the unit vector $\vec m$ and under the
angular integral we can effectively replace
\begin{equation}
m_i \; m_j \to {1\over3} \delta_{ij}.
\end{equation}
Consequently
\begin{eqnarray}
\oint \d^3 x \; \vec E^2
&=&+  
\oint \d^3 x \left\{
\nabla^2 \psi \nabla^2 \psi + {1\over3} 
( \vec  \nabla\psi) \cdot 
\nabla^2 ( \vec  \nabla\psi)
\right\}
\\
&=& +  
\oint \d^3 x 
\left\{ (\nabla^2 \psi)^2 - {1\over3}  (\nabla^2 \psi)^2 \right\}
\\
&=& +  
{2\over3} \oint \d^3 x 
\left\{ (\nabla^2 \psi)^2 \right\}.
\end{eqnarray}
Similarly
\begin{eqnarray}
\oint \d^3 x \; \vec B^2
&=&-  
\oint \d^3 x \left\{ 
(\vec m\dot\psi) \cdot \nabla^2  (\vec m\dot\psi) + 
( \vec m \cdot \vec  \nabla\dot\psi) \; 
( \vec m \cdot \vec  \nabla\dot\psi)
\right\}.
\end{eqnarray}
Again invoking spherical symmetry for $\psi$ we have
\begin{eqnarray}
\oint \d^3 x \; \vec B^2
&=&-
\oint \d^3 x \left\{ 
\dot\psi \nabla^2  \dot\psi + 
{1\over3} (\vec  \nabla\dot\psi)\cdot(\vec  \nabla\dot\psi)
\right\}
\\
&=&+ {2\over3} 
\oint \d^3 x \left\{ 
(\vec  \nabla\dot\psi)\cdot(\vec  \nabla\dot\psi)
\right\}.
\end{eqnarray}
Therefore
\begin{equation}
{\mathcal E} = \oint \d^3 x \; \rho = 
{1\over3} \oint \d^3 x \left\{ 
(\nabla^2 \psi)^2
+
(\vec  \nabla\dot\psi)\cdot(\vec  \nabla\dot\psi)
\right\},
\label{E:energy1}
\end{equation}
so the total energy is sensibly positive.  Before making further
choices regarding the potential $\psi$, consider the total momentum
\begin{equation}
\vec \wp = \oint \d^3 x \; \vec S = \oint \d^3 x \;
\left[\vec\nabla\times(\vec\nabla\times[\vec m\; \psi])\right] 
\times
\left[ \vec\nabla\times(\vec m\; \dot\psi) \right].
\end{equation}
Integration by parts implies
\begin{equation}
\vec \wp = \oint \d^3 x \left\{ 
(\vec m\; \dot\psi) \cdot \nabla^2 (\vec\nabla\times[\vec m\; \psi])
\right\}
=
-\oint \d^3 x \left\{ 
\dot\psi \vec m \cdot [\vec m \times \vec\nabla (\nabla^2  \psi)] 
\right\} = 0,
\end{equation}
so that the net momentum is zero.  For the action, an integration by
parts together with the equations of motion yields
\begin{equation}
\oint \d^4 x \; {1\over2} \left[ \vec E^2 - \vec B^2 \right] 
= 
{1\over3} \oint \d^3 x \left\{ 
(\nabla^2 \psi)^2
-
(\vec  \nabla\dot\psi)\cdot(\vec  \nabla\dot\psi) 
\right\} = 0.
\end{equation}
So we still have a zero action solution.

Up to now the potential $\psi$ has only needed to be spherically
symmetric and to satisfy the wave equation (plus some falloff
constraint at spatial infinity to allow the integration by parts). The
general solution to the wave equation in spherical symmetry is
\begin{equation}
\psi(r,t) = {1\over r} \left[ f(r+t) - f(r-t) \right].
\end{equation}
Returning to the energy ${\mathcal E}$ as given in equation
(\ref{E:energy1}), a further integration by parts, together with the
wave equation for $\psi$ yields the computationally convenient form
\begin{equation}
{\mathcal E} 
=
{1\over3} \oint \d^3 x \left\{ 
\partial_t^2 \psi \; \partial_t^2 \psi 
- \partial_t\psi \; \partial_t^3 \psi
\right\}.
\end{equation}
Whence
\begin{eqnarray}
{\mathcal E} 
&=&
 4\pi \int_0^\infty \d r \Big\{ 
(\partial_t^2[ f(r+t) - f(r-t) ] )^2 
\\
&&\qquad \qquad 
- 
(\partial_t[ f(r+t) - f(r-t) ] ) \; (\partial_t^3[ f(r+t) - f(r-t) ] )
\Big\}
\nonumber
\\
&=&
 4\pi \int_0^\infty \d r \Big\{ 
(\partial_r^2[ f(r+t) - f(r-t) ] )^2 
\\
&&\qquad\qquad 
- 
(\partial_r[ f(r+t) + f(r-t) ] ) \; (\partial_r^3[ f(r+t) + f(r-t) ] )
\Big\}
\nonumber
\\
&=&
 4\pi \int_0^\infty \d r \Big\{ 
(\partial_r^2[ f(r+t) - f(r-t) ] )^2 
\\
&&\qquad\qquad 
+ 
(\partial_r^2[ f(r+t) + f(r-t) ] )^2
\Big\}
\nonumber
\\
&=&
8\pi \int_0^\infty \d r \left\{ 
[\partial_r^2 f(r+t) ]^2 + 
[\partial_r^2 f(r-t) ]^2
\right\}
\\
&=&
8\pi \int_{-\infty}^{+\infty} \d s \; 
[\partial_s^2 f(s) ]^2.
\end{eqnarray}
Which verifies that the energy is constant in a model-independent
manner. Furthermore if $f(s)$ is smooth and satisfies suitable falloff
conditions at $s\to\pm\infty$ then the wavelet will be nonsingular and
of finite energy.

To obtain a specific example it only remains to finish the complete
specification of the potential $\psi(r,t)$. One particularly simple
choice is to take one of the real scalar wavelets of the previous
section $\psi\to\phi_{1,2}$, with $\zeta^a = ||\zeta|| \; V^a = a\;
V^a$. Since the electric and magnetic fields are now specified in
terms of derivatives of a smooth bounded function, the electromagnetic
field is similarly smooth and bounded.  For the energy, write it in
the form
\begin{equation}
{\mathcal E} =
{1\over3} \oint \d^3 x \left\{ 
\psi \; \partial_t^4 \psi - \partial_t\psi \; \partial_t^3 \psi
\right\}.
\end{equation}
Whence, for either $\psi\to\phi_{1,2}$, an integration carried out at
arbitrary $t$ yields the time-independent quantity
\begin{equation}
{\mathcal E}_{1,2} = {1\over2} \;{\phi_0^2\over a}.
\end{equation}
The calculation has for convenience been carried out in the
zero-momentum frame of the wavelet. In a general Lorentz frame we
would have
\begin{equation}
P^a_{1,2}  = {1\over2}\; {\phi_0^2\over a} \;\; V^a 
=  {1\over2}\; {\phi_0^2\over a^2} \;\; \zeta^a.
\end{equation}
Thus this field configuration, as for the scalar case, is Lorentz
covariant, bounded, finite energy and zero action. The vector nature
of the Maxwell field has added technical complications, but there is
no real change in basic principles.

\noindent{\underline{\bf Aside:}}
In closing this section, I should mention one other wavelet that is
particularly attractive. If one makes use of the pseudo-differential
operator $(\nabla^2)^{-1/2}$ one could write
\begin{equation}
\psi = (\nabla^2)^{-1/2} \; \phi_{1,2} = 
{1\over\Gamma(-1/2)} \int_0^\infty {\d t\over t^{3/2}} 
\exp[-t \nabla^2] \; \phi_{1,2}.
\end{equation}
With this choice of $\psi$ the energy integral simplifies
\begin{equation}
{\mathcal E}_{1,2} = 
{1\over3} \oint \d^3 x \left\{ 
(\partial_t\phi_{1,2})^2 + (\nabla \phi_{1,2})^2 
\right\}
= {1\over 6} \; \pi^2 \; |\phi_0^2| \; a.
\end{equation}
The price paid for making the energy look simple is that the electric
and magnetic fields are much more complicated to calculate.

\noindent{\underline{\bf Comment:}} 
Because the optics and engineering literature generally does not use
manifestly Lorentz covariant notation, it can be very time consuming
to calculate the 4-momentum of a specific pulse. Indeed only very
recently~\cite{Lekner} has Lekner provided specific and explicit
computations of both ${\mathcal E}$ and $\vec\wp$ (as well as the
angular momentum) for a pulse similar to that considered above. As
expected (once one has the covariant perspective advocated in this
article) $||\vec\wp|| < {\mathcal E}$, indicating the existence of a
zero-momentum frame for the pulses of this type~\cite{Lekner}.

\section{Yang-Mills field}

Once we have the Maxwell wavelet above, a Yang--Mills wavelet is
straightforward, indeed trivial. Let $\mathbf{\Lambda}$ be any
constant matrix in the center of the gauge group and set $\mathbf{A}^a
= A^a \; \mathbf{\Lambda}$. The construction is so simple that there
is really nothing extra beyond the Maxwell wavelet considered above.

\section{Null $\zeta$}
\label{S:null}

Let us now return to the original scalar complex wavelet.  Suppose the
4-vector $\zeta$ is null. Then because the numerator vanishes
identically the original definition above gives $\phi\equiv 0$. We
should at a minimum change our field definition to read
\begin{equation}
\phi(x) = - 
{\psi_0
\over
\eta_{ab} \;[x^a-x_0^a-i\zeta^a] \; [x^b-x_0^b-i \zeta^b]
}.
\end{equation}
Then without loss of generality we go into the frame
\begin{equation}
\zeta^a = (a,0,0,a),
\end{equation}
and then
\begin{equation}
\phi(x) = 
- {\psi_0
\over
[t-ia]^2 - x^2-y^2-[z-ia]^2
}.
\end{equation}
That is
\begin{equation}
\phi(x) = 
{\psi_0
\over
r^2-t^2+2ia(t-z)
}.
\end{equation}
Note
\begin{equation}
|\phi(x)| = 
{\psi_0
\over
\sqrt{(r^2-t^2)^2 +4 a^2 (t-z)^2 } 
}.
\end{equation}
The denominator now vanishes when $z-t$ and $x=y=0$, so that the field
is divergent on the beam axis. Suppose we write $R = \sqrt{x^2+y^2}$
then
\begin{equation}
\phi(x) = 
{\psi_0
\over
R^2+ z^2-t^2+2ia(t-z)
}.
\end{equation}
So we see that the field drops off as $1/R^2$ as we move away from the
beam axis. [And more critically, the field blows up as $1/R^2$ as we
approach the beam axis.] Attempts at calculating the energy and action
now lead to divergent integrals.  In other words, despite the fact
that it still solves the wave equation, for null $\zeta$ this is not a
particularly useful field configuration.

\section{Spacelike $\zeta$}
\label{S:spacelike}

Foe the complex scalar wavelet, suppose the 4-vector $\zeta$ is
spacelike. Then without loss of generality we go into the infinite
velocity frame where
\begin{equation}
\zeta^a = (0,0,0,a),
\end{equation}
and then
\begin{equation}
\phi(x) = 
{\phi_0 \; a^2
\over
t^2 - x^2-y^2-[z-ia]^2
}.
\end{equation}
That is
\begin{equation}
\phi(x) = 
-{\phi_0 \; a^2
\over
r^2-t^2-a^2-2iaz
}.
\end{equation}
Note
\begin{equation}
|\phi(x)| = 
{\phi_0 \; a^2
\over
\sqrt{(r^2-t^2-a^2)^2 +4 a^2 z^2 } 
}.
\end{equation}
The denominator now vanishes when $z=0$ and $x^2+y^2=a^2+t^2$.  That
is, the field is divergent on a time-dependent circle orthogonal to
the beam axis. There is a short distance singularity as one approaches
this circle, and the energy and action integrals diverge.  In other
words, despite the fact that it still solves the wave equation, for
spacelike $\zeta$ this is not a useful field configuration.

\section{Discussion}

The physical wavelet discussed in this article is important because it
represents a qualitatively different extended field configuration of a
type not normally encountered in particle physics. The wavelet is
neither a soliton, nor an instanton, nor a sphaleron though it
shares properties with all three of these extended objects:
\begin{itemize}

\item Like the soliton it lives in physical time (Minkowski space, not
  Euclidean space), and possesses a well-defined 4-velocity.

\item Like the instanton it ``dies away'' in the infinite past and future.
  
\item Like the instanton it possesses a continuously adjustable scale
  parameter.

\item Like the sphaleron it is unstable to dispersal.
\end{itemize}
Because the wavelet fields are bounded and finite energy, wavelet
configurations {\emph{will}} be classically excited at any finite
temperature. Because the wavelet configuration has zero action,
arbitrarily complicated combinations of these physical wavelets can be
added to the field configurations appearing in Feynman's path integral
without modifying the phase --- quantum mechanically there is no
``cost'' in adding these configurations to the Lorentzian path integral
and they {\emph{will}} contribute.

Other ``localized waves'' might be interesting in specific
applications but the particular example discussed in this article is
important because of its extreme simplicity and pleasant behaviour.

\acknowledgments

I wish to thank John Lekner for stimulating my interest in these
issues. I also wish to thank Damien Martin for bringing the whole
Lorenz/Lorentz issue to my attention.



\begin{thebibliography}{99}

\bibitem{zed}
R. W. Ziolkowski, 
``Exact solutions of the wave equation with complex source locations'',
J. Math. Phys. {\bf 26} (1985) 861--863.

\bibitem{tippet}
M. K. Tippet and R. W. Ziolkowski, 
``A bidirectional wave transformation of the cold plasma equations'',
J. Math. Phys. {\bf 32} (1991) 488--492.

\bibitem{LG}
J. Lu and J. F. Greenleaf,
``Nondiffracting X waves --- exact solutions to free-space 
scalar wave equation and their finite aperture realizations'',
IEEE Transactions on ultrasonics, ferroelectrics, and frequency control, 
{\bf 39} (1992) 19--31.

\bibitem{LZG}
J. Lu, H. Zou, and J. F. Greenleaf,
``A new approach to obtain limited diffraction beams'',
IEEE Transactions on ultrasonics, ferroelectrics, and frequency control, 
{\bf 42} (1995) 850--853.

\bibitem{Kaiser}
G.~Kaiser,
``Physical wavelets and their sources: Real physics in complex spacetime,''
arXiv:math-ph/0303027.

\bibitem{Lekner}
J. Lekner,
``Electromagnetic pulses which have a zero momentum frame'',
submitted to J. Opt. A, 21 November 2002; arXiv:physics/0304022.

\bibitem{key} Field configurations similar to the one considered in
  this article may variously be encountered under names such as:
  ``localized waves'', ``focus wave modes'', ``pulses'', ``X-waves'',
  ``limited diffraction beams'', ``wavelets'', and ``physical
  wavelets''.
  
\bibitem{Lorenz} The Lorenz gauge was apparently first used by the
  Danish physicist Ludwig Lorenz (1829-1891), though it is commonly
  misattributed to the Dutch physicist Hendrik Antoon Lorentz
  (1853-1928).
\\
L.  Lorenz, 
``On the Identity of the Vibrations of Light with Electrical Currents'',
Philos. Mag. {\bf 34} (1867) 287-301. 
\\
J. van Bladel, ``Lorenz or Lorentz?'', 
IEEE Antennas Prop. Mag. {\bf 33} (1991) 69. 


\end{thebibliography}
\end{document}